\documentclass[twocolumn,showpacs,showkeys,superscriptaddress,prb]{revtex4}
\usepackage{graphicx}
\usepackage{dcolumn} 

\newcommand{\st}{SrTiO$_3$}

\newcommand{\bt}{BaTiO$_3$}
\newcommand{\btd}{BaTiO$_{3-\delta}$}

\newcommand{\std}{SrTiO$_{3-\delta}$}

\begin{document}

\title{Doping and temperature dependent optical properties of reduced \btd}
\author{J. Hwang}
 \email{jhwang@pusan.ac.kr}
 \affiliation{Department of Physics and Astronomy, McMaster University, Hamilton, ON L8S 4M1, Canada}
 \affiliation{Department of Physics, Pusan National University, Busan 609-735, Republic of Korea}
\author{T. Kolodiazhnyi}
 \email{kolodiazhnyi.taras@nims.go.jp}
 \affiliation{National Institute for Materials Science, 1-1 Namiki, Tsukuba, Ibaraki 305-0044, Japan}
\author{J. Yang}
 \affiliation{Department of Physics and Astronomy, McMaster University, Hamilton, ON L8S 4M1, Canada}
 \affiliation{Tianjin Key Laboratory of Composite and Functional Materials, School of Materials Science and Engineering, Tianjin University, Tianjin, 300072, PR China}
\author{M. Couillard}
 \affiliation{Canadian Centre for Electron Microscopy, McMaster University, Hamilton, ON L8S 4M1, Canada}

\date{\today}

\begin{abstract}
We report on optical properties of reduced \btd\ at different doping levels including insulating and metallic samples. In all the samples, including metallic one, we observe structural phase transitions from the changes of the infrared active phonon modes. Metallic ground state is confirmed by the Drude-type low-frequency optical reflectance. Similar to \std\ we find that the mid-infrared absorption band in \btd\ appears and grows with an increase in the oxygen vacancy concentration. Upon decrease in temperature from 300 K, the mid-infrared band shifts slightly to higher frequency and evolves into two bands: the existing band and a new and smaller band at lower frequency. The appearance of the new and smaller band seems to be correlated with the structural phase transitions.

\end{abstract}

\pacs{78.20.-e, 72.20.-i, 78.20.Ci}

\maketitle

\section{Introduction}

Recent attention to \bt\ and \st\ perovskites as possible hosts for tunable superconductors\cite{caviglia08}, thermoelectric devices\cite{ohta07}, and non-destructive tunnel resistance ferroelectric memory\cite{garcia09} calls for better understanding of their electronic properties. Different strength of electron-phonon coupling in n-type \st\ and \bt\ have been invoked to classify these materials into the \emph{large} and \emph{small} polaron categories, respectively.\cite{verbist92,eagles96,lenjer02,kolodiazhnyi06} The origin of the different nature of the electron-phonon interactions in these two compounds is not clear, however, especially considering very similar lattice constants and identical valence electronic structure.

Although recent studies suggest that the two perovskites posses more similarities than previously thought, significant differences exist in the magnetic and electronic properties of \st\ and \bt, especially in metallic phase. For example, metallic \st\ shows nearly temperature-independent Pauli paramagnetism as well as $T$-independent Hall coefficient. Metallic \bt, on the other hand, shows correlation among strongly $T$-dependent magnetic susceptibility, Hall and Seebeck coefficients.\cite{kolodiazhnyi08} Certain problems with the classical small polaron interpretation of the charge transport in n-type \bt\ have been addressed by one of the authors\cite{kolodiazhnyi03} who has also explained insulator-metal-transition (IMT) in \btd\ in terms of a scaled Mott criterion.\cite{kolodiazhnyi08,edwards:1978}

\begin{figure}[t!]
        \begin{center}
         \vspace*{-0.0 cm}%
               \leavevmode
                \includegraphics[origin=c, angle=0, width=5 cm, clip]{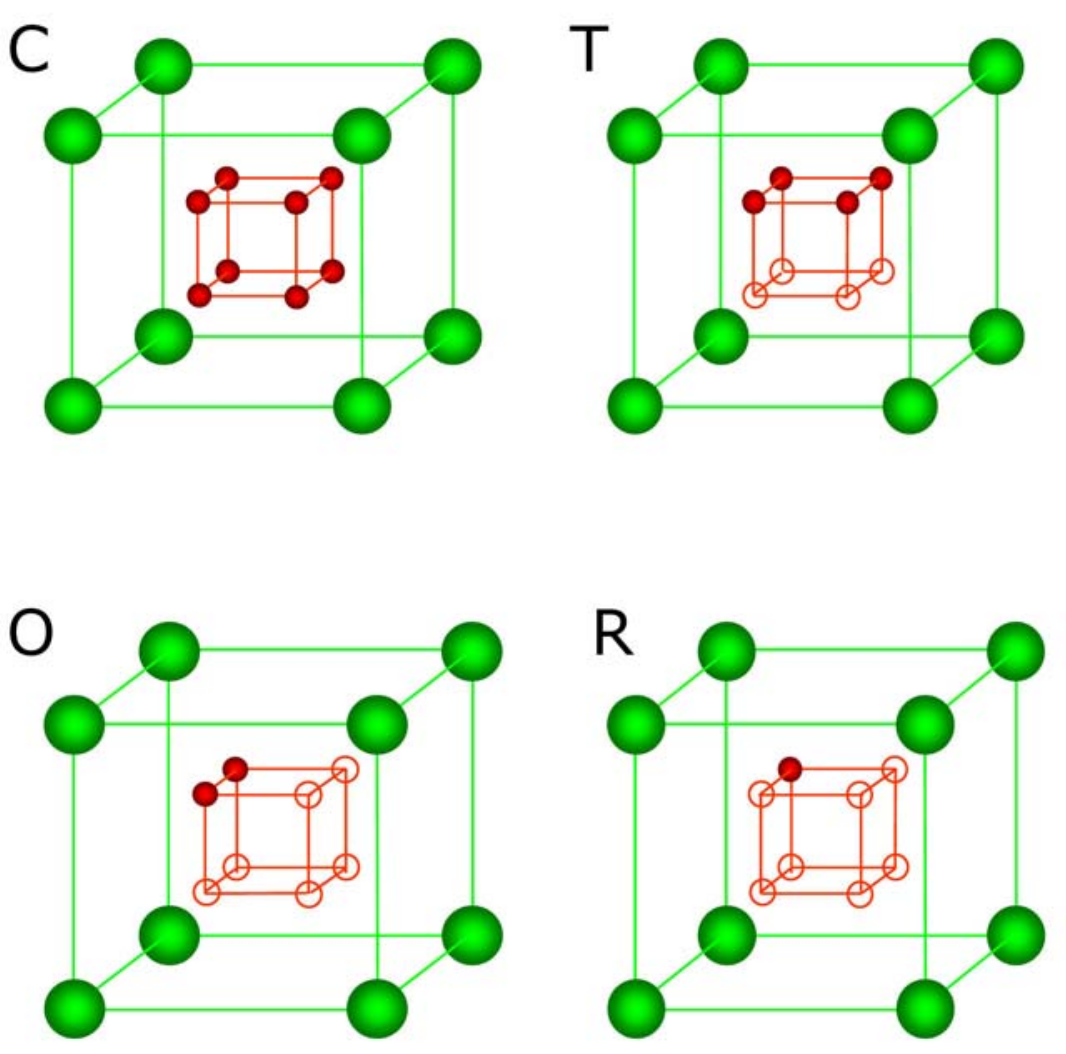}
         \vspace*{-0.0 cm}%
        \end{center}
  \caption{(Color online) Phase transformations in \bt\ according to order-disorder model of Comes et al. \cite{comes70}. C, T, O, and R designate cubic, tetragonal, orthorhombic and rhombohedral phases,
  respectively. Large (green) and small (red) spheres designate Ba and Ti ions, respectively. Oxygen octahedra are not shown. Off-center displacements of Ti ion
  along $[$111$]$ body diagonals are exaggerated for clarity. Note the change in the positional degeneracy of the Ti ion in different phases.
  The position of Ti ion is 8-, 4-, 2- and 0-fold degenerate in the C, T, O, and R phases, respectively.}
  \label{fig1}
\end{figure}

Strong coupling of charge carriers with the high-energy longitudinal optical (LO) phonons has been advocated by classical small polaron theories to explain thermally activated electron mobility.\cite{holstein59} In contrast, piezoresistance measurements on \bt, \st, KTaO$_3$ and KTa$_{1-x}$Nb$_x$O$_3$ reveal that conduction electrons interact most strongly with the low-energy (soft) transverse optical (TO) phonons, with \bt\ demonstrating the highest piezoresistance effect among the crystals studied.\cite{wemple66} Another channel of electron scattering in \bt\ that, until now, has been completely ignored is related to the positional disorder of Ti ions. Comes et al.\cite{comes70} proposed an order-disorder model of the phase transitions in \bt\ based on the 8-fold off-center positional degeneracy of Ti ions (Figure \ref{fig1}).

%
%

Recent findings provided by terahertz and infrared (IR) data on tetragonal \bt\ indicate that the characteristic Debye frequency of the central (relaxation) mode associated with the off-center Ti dynamics is of the order of 100 cm$^{-1}$ which is comparable with the eigenfrequency of the low-energy (soft) phonons.\cite{hlinka08} In view of extremely large amplitude of Ti off-center displacements (e.g., 0.19 \r{A}) one may expect strong dynamic perturbation of the conduction band width at the onset of the Ti disorder.\cite{ravel98} In fact, disorder in the form of microscopic clusters of orthorhombic Ti off-center displacements exists already in rhombohedral phase as revealed by Mn$^{4+}$ electron paramagnetic resonance (EPR).\cite{muller86,volkel07} Evidence of both order-disorder and displacive character of the phase transitions in \bt\ has been found by variety of techniques including nuclear magnetic resonance\cite{zalar03} (NMR), extended x-ray-absorption fine structure (EXAFS), x-ray-absorption near-edge structure\cite{ravel98} (XANES), Raman\cite{vogt82} and IR\cite{gervais84} spectroscopy.  It becomes obvious, therefore, that both resonant (optical phonons) and non-resonant (relaxation) modes have to be considered in the problem of electron-lattice interaction in \bt.

Significant insight into the interaction of itinerant electrons with the lattice degrees of freedom is provided by a wide band optical spectroscopy. Most of the literature so far has been dedicated to n-type \st,\cite{calvani93,gervais93,crandles99,bi06,mechelen08} and only a limited number of papers have addressed optical properties of n-type \bt.\cite{berglund67,zhao00} Crandles {\it et al.} studied highly reduced \std\ by using optical spectroscopy.\cite{crandles99} They observed broadening of the phonon modes with an increase in the oxygen vacancy,
$V_O$, density as well as $V_O$-induced mid-infrared band whose strength scales with carrier concentration.\cite{crandles99} The most reduced sample, SrTiO$_{2.72}$, displayed a localization-modified Drude optical conductivity that was attributed to the dynamic disorder introduced by the oxygen vacancies.\cite{crandles99} In the most recent studies, enhancement of effective electron mass by a factor of 2-3 has been deduced from a narrow Drude peak and mid-infrared (MIR) absorption band in the Nb-doped \st. \cite{mechelen08} It is worth mentioning that the origin of the MIR band in cuprates, nickelates and titanates remains controversial. MIR absorption band in n-type \st\ and \bt\ has been explained by different authors in terms of the disorder-assisted Ti 3$d$ intraband scattering,\cite{calvani93} photoionization of the F centers,\cite{crandles99,berglund67} multiphonon absorption,\cite{mechelen08} and optical excitations of small/large polarons.\cite{verbist92,bi06}

In the paper we report on correlations between electron and lattice dynamics in \btd\ single crystals as evidenced by significant changes in the temperature and doping dependence of optical reflectivity and conductivity. IR data confirm that both insulating and metallic samples undergo crystallographic phase transitions, thus supporting recent findings of `ferroelectric metal' ground state in \btd.\cite{kolodiazhnyi10} The most remarkable effect of electron doping is found in colossal renormalization of the soft TO mode in \btd. In contrast to n-type \st\ where the soft mode energy increases with the free electron density,\cite{gervais93,mechelen08,bauerle80} the energy of the TO soft mode in the rhombohedral phase of \bt\ shows pronounced drop upon increase in electron concentration. In addition to the Drude part, optical conductivity of metallic \btd\ reveals that the soft TO mode splits into low- and high-frequency components. The energy of the high-frequency component recovers to that of undoped \bt.

\section{Experimental methods}

Details of the single crystal preparation as well as dc resistivity, Hall, thermoelectric power, and magnetic measurements of \btd\ have been already reported in Ref. \cite{kolodiazhnyi08} For optical studies we selected four \btd\ crystals with charge carrier concentrations of $n < 1 \times 10^{12}$ (undoped), 3.9$\times 10^{17}$, 3.1$\times 10^{19}$, and 2.0$\times 10^{20}$ cm$^{-3}$ designated hereafter as B1, B2, B3 and B4, respectively. For undoped and lightly reduced \btd\ with $n  = 3.9 \times 10 ^{17}$ cm$^{-3}$ it was possible to identify single domain crystals in the tetragonal phase (270 K $<T<$ 403 K) by a polarizing microscope. For more heavily reduced (non-transparent) samples, we took advantage of anisotropic resistivity in tetragonal phase\cite{berglund66} for selection of the single domain crystals for optical measurements.

Reflectance data were obtained in a spectral range from 50 cm$^{-1}$ to 40,000 cm$^{-1}$ by using a commercial Fourier transform infrared (FTIR)-type spectrometer, Bruker 66v/S with spectral range extension packages (both far infrared and ultraviolet). Optical measurements were performed at selected temperatures in the 20 - 300 K range using a continuous flow liquid $^4$He cryostat made by R. G. Hansen \& Associate. For accurate reflectance measurement an {\it in-situ} gold evaporation method\cite{homes93} was employed. In this method gold (below 20,000 cm$^{-1}$) and aluminium (above 20,000 cm$^{-1}$) films (typical thickness $\geq$ 200 nm) evaporated on the sample were used as the references for the reflectivity. Absolute reflectance was obtained by multiplying the absolute reflectance of the coating material to the measured relative reflectance with respect to the coating material. An automatic temperature-controlled reflectance measurement system ensured reproducible and reliable measurement of reflectance data.

For reflectance data above 40,000 cm$^{-1}$ we measured undoped BaTiO$_3$ up to 175 eV by using electron energy loss spectroscopy (EELS) technique.\cite{egerton09} EELS characterization was carried out on a JEOL 2010F transmission electron microscope (TEM) with a Schottky field emission gun operated at 200 keV. The microscope is equipped with a Gatan Tridiem energy filter for EELS measurements. Because in an electron energy-loss spectroscopy one measures the dielectric response of the sample, it is possible to extract the real and imaginary part of the dielectric functions, and therefore the reflectance of materials. A Kramers-Kronig (KK) analysis \cite{egerton96} was performed on a single scattering distribution (SSD) extracted from the EELS spectrum using deconvolution routines available in the DigitalMicrograph environment. The SSD is proportional to the energy-loss function $Im[-1/\varepsilon(\omega)]$, where $\varepsilon(\omega)$ is the complex dielectric function. Extensions to the spectra for energy loss above 200 eV ensured an accurate energy integration in the Kramers-Kronig transformations, and corrections to take into account the collection angle and the surface contributions were also included. Finally, the dielectric function was normalized using the optical refractive index. The reflectance was then obtained by using the Fresnel formula, $R(\omega)=|(\sqrt{\varepsilon(\omega)}-1)/(\sqrt{\varepsilon(\omega)}+1)|^2$. The EELS data were used for extrapolation of the optical reflectance to the high frequency side. Optical conductivity was obtained by the KK transformation.\cite{wooten72,aspness85} For KK analysis one has to extrapolate the measured reflectance into both low ($\omega$=0) and high ($\omega=\infty$) frequency sides. For the $\omega$=0 extrapolation of conducting samples we have used the measured dc resistivity, $\rho$, assuming Hagen-Ruben relationship, $R=1-2\sqrt{\omega \rho/2 \pi} $. For the $\omega$=0 extrapolation of insulating sample we have used a low-frequency dielectric constant data available from literature.\cite{merz49} For the $\omega=\infty$ extrapolation we have applied $R(\omega) \propto \omega^{-4}$ relation above the maximum measured frequency of 175 eV. A detailed procedure for KK analysis of optical data can be found, for example, in Ref. \cite{hwang07}.

\section{Results}

    \subsection{dc resistivity}
%
%
\begin{figure}[]
        \begin{center}
         \vspace*{-2.0 cm}%
               \leavevmode
                \includegraphics[origin=c, angle=0, width=8 cm, clip]{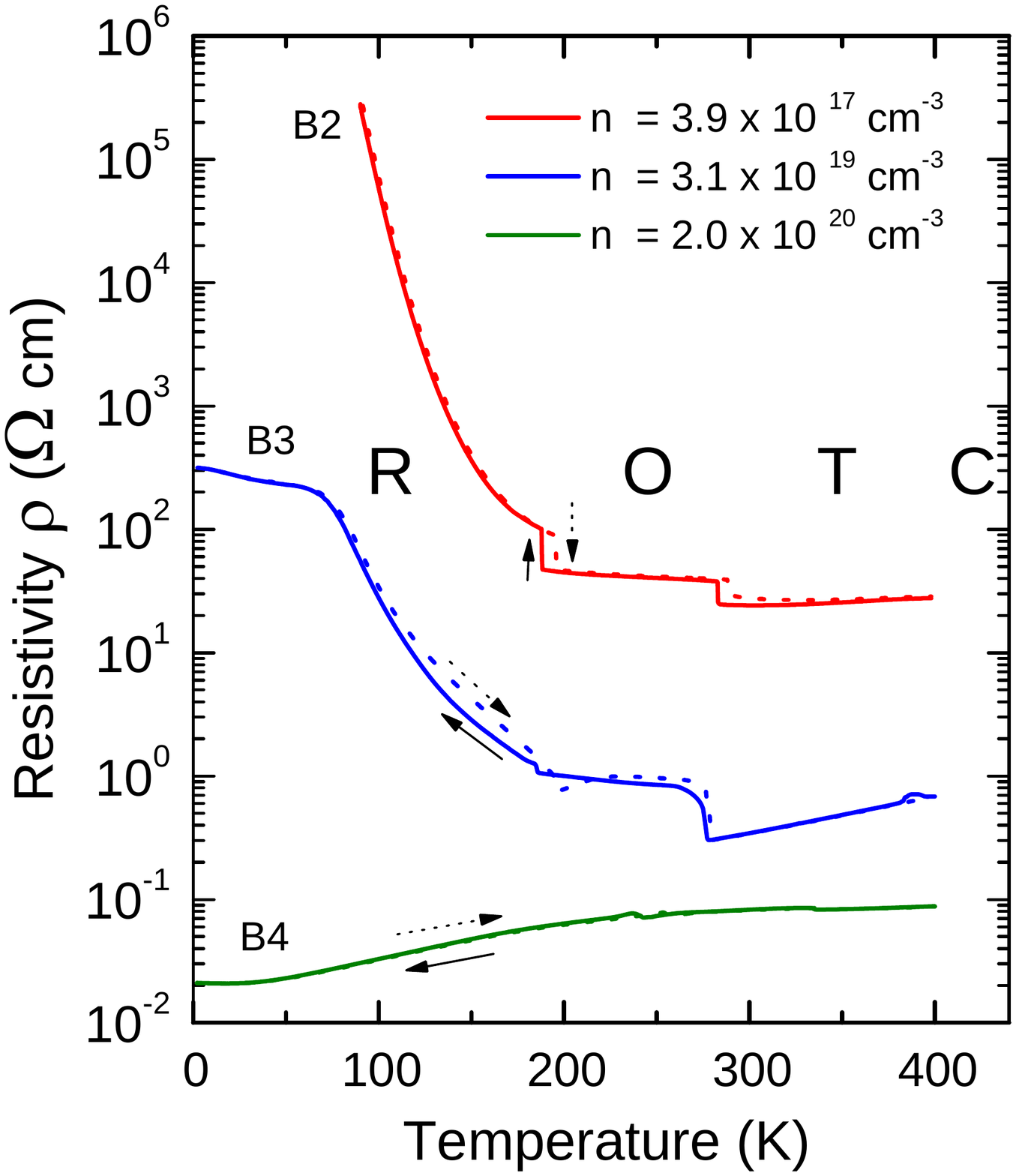}
         \vspace*{-2.5cm}%
        \end{center}
\caption{(color online) Semilog plot of dc resistivity of three \btd\ samples B2, B3 and B4 with electron concentration of 3.9$\times 10 ^{17}$, 3.1$\times 10^{19}$, and 2.0$\times 10^{20}$ cm$^{-3}$, respectively. R, O, T, and C stand for rhombohedral, orthorhombic, tetragonal, and cubic phases, respectively.}
\label{fig2}
\end{figure}

\begin{figure}[t!]
        \begin{center}
         \vspace*{-1.0 cm}%
               \leavevmode
                \includegraphics[origin=c, angle=0, width=8 cm, clip]{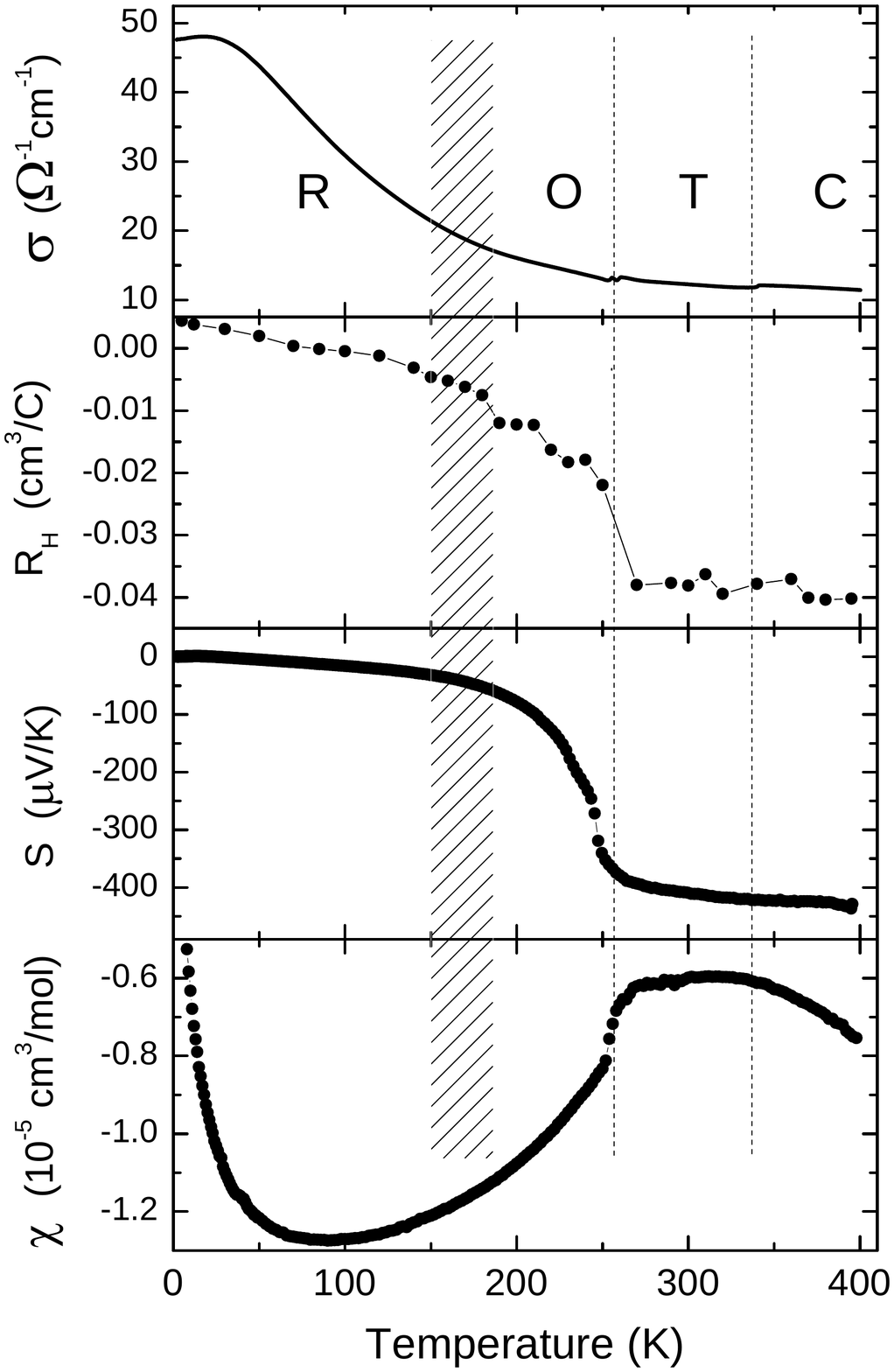}
         \vspace*{-1.0cm}%
        \end{center}
\caption{dc conductivity, $\sigma$, Hall coefficient, $R_H$, Seebeck coefficient, $S$, and magnetic susceptibility, $\chi$ of the metallic \btd\ with n = 2.0$\times$10$^{20}$ cm$^{-3}$ adopted from Ref. \cite{kolodiazhnyi08}. The vertical lines indicate phase transition temperatures determined by differential scanning calorimetry. The boundary of the lowest phase transition temperature was not clearly identified and is shown as a broad hatched vertical line.} \label{fig3}
\end{figure}

Figure \ref{fig2} shows dc resistivity, $\rho$, of the B2, B3 and B4 crystals. The data were taken on both cooling (solid line) and heating (dotted line). One can notice hysteresis in the $\rho$ vs. $T$ dependence at the structural phase transitions. The electron concentrations, $n$, were obtained from $R_H$ measurements at 400 K. The B2 sample ($n = 3.9 \times 10 ^{17}$ cm$^{-3}$) shows insulating behavior in the R phase with an activation energy of ca. 0.1 $\pm$ 0.01 eV. Heavily reduced B4 sample ($n = 2.0 \times 10^{20}$ cm$^{-3}$) is metallic with resistivity decreasing on cooling. The B3 sample with $n = 3.1 \times 10^{19}$ cm$^{-3}$ shows almost temperature independent resistivity
below 80 K and is located on the insulating side of the metal-insulator transition. In a separate study we confirm that anomalous steps in the dc resistivity (Figure \ref{fig1}) correspond to the phase transition temperatures determined by differential scanning calorimetry.\cite{kolodiazhnyi10} Doping with oxygen vacancies causes significant reduction in the phase transition temperature in agreement with Ref. \cite{hardtl72} The stability of the ferroelectric (FE) domain pattern can be confirmed by recovery of electrical resistivity to the same values upon thermal cycling in the 2--400 K range. The three-fold orbital degeneracy of Ti t$_{2g}$ electrons that exists in the cubic phase is removed upon entering the T phase. This is reflected in significant anisotropy of electron transport in the T phase.\cite{berglund66,kolodiazhnyi08}

Figure \ref{fig3} shows temperature dependence of dc conductivity, $\sigma$, Hall coefficient, $R_H$, Seebeck coefficient, $S$, and molar magnetic susceptibility, $\chi$, of metallic B4 sample. A two-band conduction model with a narrow energy gap has been advocated by one of the authors in an attempt to explain strong temperature dependence of $R_H$, $S$ and $\chi$ of metallic \bt.\cite{kolodiazhnyi08} This model has yet to be confirmed or rejected by the first-principles calculations of \bt\ in the low-symmetry phases. An alternative explanation of the anomalous magnetic and thermoelectric behavior of metallic \btd\ shown in Fig. \ref{fig3} may require assumption of the temperature-dependent electron effective mass, $m^{\ast}(T)$. Strong renormalization of effective mass can be attributed to the onset of the Ti off-center positional disorder which would cause significant perturbation of the conduction band edge. Enhancement of effective mass is associated with an increase in the density of states at the Fermi level and slowing down of the Fermi velocity. Since renormalization of effective mass is expected to scale linearly with the renormalization of the relaxation time, $\tau$$^{\ast}$, i.e.,

\begin{equation}
m^\ast/m = \tau^\ast/\tau, \label{eq1}
\end{equation}
the electron mobility $\mu =$ e$\tau^{\ast}/m^{\ast}$ is not affected, and one may not observe significant changes in the dc conductivity. The Drude component of the optical conductivity,
\begin{equation}
\sigma_1(\omega) = \sigma_0 \left(\frac{1}{1+\omega^2\tau^{\ast2}}\right), \label{eq2}
\end{equation}
$\sigma_1(\omega)$, on the other hand, may undergo noticeable changes provided that the $\tau^{\ast}(T)$ enhancement is relatively steep. In Eq. \ref{eq2} $\sigma_0 = n e^2 \tau^{\ast}/m^{\ast}$ is the dc conductivity, $\omega$ is the angular frequency and $e$ is the elementary charge. If this scenario holds for metallic \btd, the most noticeable changes in the optical data would be expected in the temperature range of 200--275 K where material undergoes transformation from orthorhombic to tetragonal phase (Fig. \ref{fig3}).

\subsection{Optical reflectance}
%
\begin{figure}[t!]
        \begin{center}
         \vspace*{-0.4 cm}%
               \leavevmode
                \includegraphics[origin=c, angle=0, width=8 cm, clip]{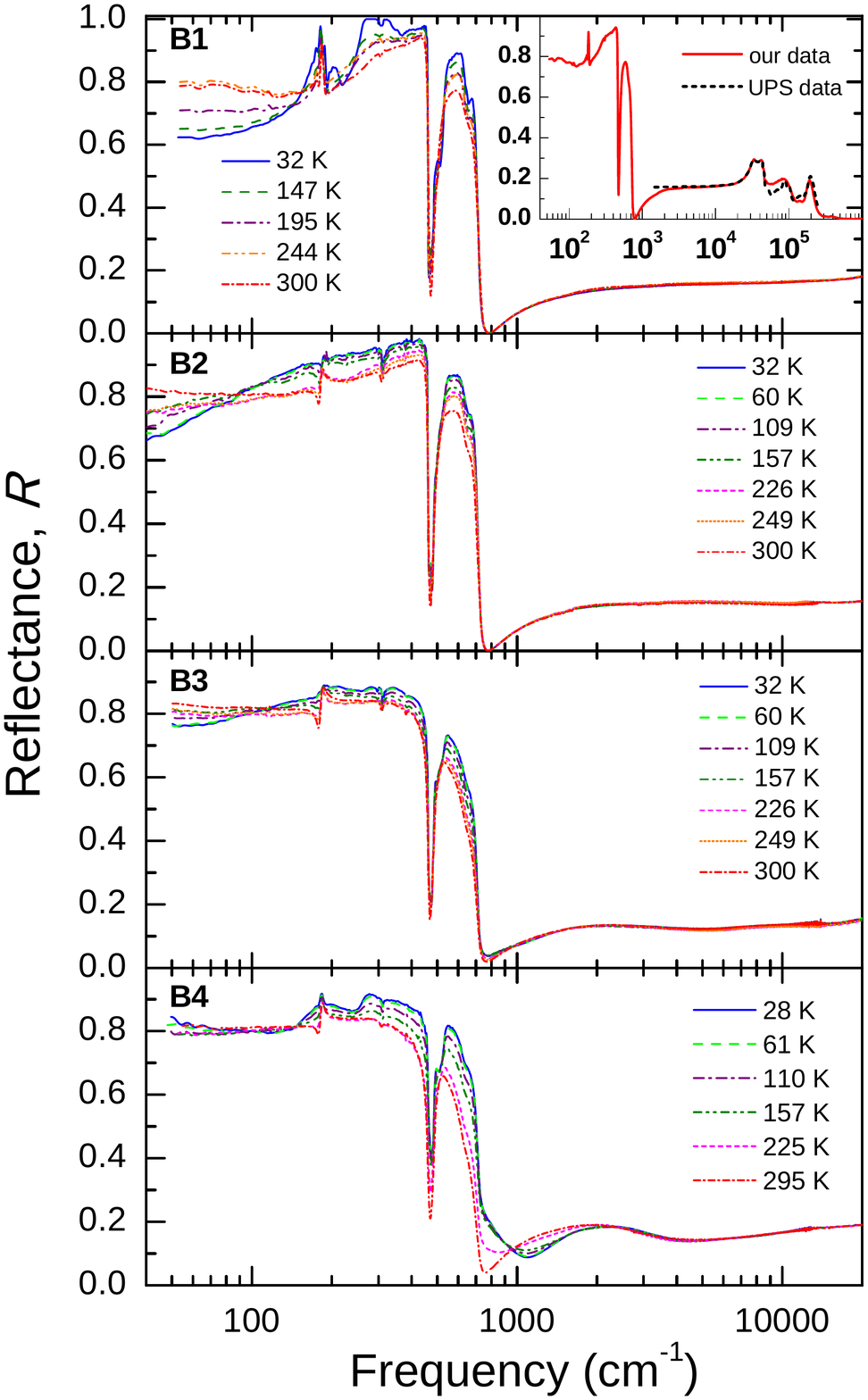}
         \vspace*{-0.7cm}%
        \end{center}
\caption{(color online) Measured ab-plane reflectance of the four \btd\ samples at various temperatures. The inset in the B1 panel shows room temperature reflectance data for undoped BaTiO$_3$ up to 1,400,000 cm$^{-1}$ ($\simeq$ 175 eV). The high frequency reflectance above 40,000 cm$^{-1}$ is obtained from the electron energy loss spectroscopy. The dash line in the inset shows reflectance derived from the ultraviolet photoelectron spectroscopy (UPS) adopted from work by B\"{a}uerle {\it et al.}\cite{bauerle78}. Panels B1--B4 designate the samples' labels.} \label{fig4}
\end{figure}

Figure \ref{fig4} shows ab-plane reflectance of the \btd\ samples with different electron concentrations at various temperatures. The inset in the B1 panel shows an extended room temperature reflectance of undoped \bt\ up to 1.4$\times 10^{6}$ cm$^{-1}$ ($\simeq$ 175 eV) obtained from the EELS measurements. Our EELS data are in good agreement with reflectance derived from the ultraviolet photoelectron spectroscopy (UPS)\cite{bauerle78} also shown in the inset of panel B1.

Insulating \bt\ shows temperature-dependent reflectance only below 700 cm$^{-1}$ (Figure \ref{fig4}, panel B1). The low-frequency ab-plane reflectance $R$ = 0.8 of this sample at 300 K is roughly 20 \% larger than that of tetragonal (c-axis) reflectance reported in Ref. \cite{hlinka08} However, it is still lower than the ab-plane reflectance of $R \approx$ 0.85 reported by Spitzer et al.\cite{spitzer62} We cannot exclude, therefore, that the ab-plane reflectance data of our crystal have an admixture of the c-plane reflectance. At $T$ = 32 K the low-frequency reflectance of the insulating \bt\ drops to 0.62 which corresponds to the $\varepsilon \approx$ 72, in excellent agreement with
capacitance data at 1 kHz (not shown). Overall, the low-frequency reflectance of insulating, B1, and semiconducting, B2 and B3, samples increases with $T$ in the 30--300 K range which is most likely due to the temperature-dependent phonon contribution to the dielectric constant. Doping gives pronounced effects on the low-frequency reflectivity of \btd. First, the $T$-dependence of the low-frequency reflectnace decreases with doping. Second, the low-$T$ reflectivity below 200 cm$^{-1}$ increases with doping in both semiconducting, B2 and B3, and metallic B4 samples. The low-$T$ conductivity of the B2 and B3 samples is still too low; therefore, we rule out possible Drude contribution to the low-frequency dielectric constant.

Temperature-dependent region of reflectance in metallic \btd\ extends to 2500 cm$^{-1}$ (Figure \ref{fig4}, sample B4) which is well beyond the phonon contribution. Low-frequency reflectance of B4 sample is the highest at 28 K in contrast to other samples studied. This temperature dependence of the low frequency reflectance agrees with the dc resistivity data (see Figure\ref{fig2}) and is associated with the Drude contribution from itinerant electrons. Strong reflectance dips at 470 and 750 cm$^{-1}$ from the IR active phonons become shallower due to the screening by the free charge carriers. Metallic \btd\ also reveals very strong temperature and frequency dependence of reflectance valley in the 750--1200 cm$^{-1}$ range which, in our opinion, also originates from the temperature dependent Drude contribution.

Another noticeable effect of doping is found in the mid IR range of 2500--9000 cm$^{-1}$ where a broad reflectance valley develops in the samples with increasing electron concentration (see Figure \ref{fig4}, samples B3 and B4).

    \subsection{Optical conductivity}

%
%
\begin{figure}[t!]
        \begin{center}
         \vspace*{-0.4 cm}%
               \leavevmode
                \includegraphics[origin=c, angle=0, width=8 cm, clip]{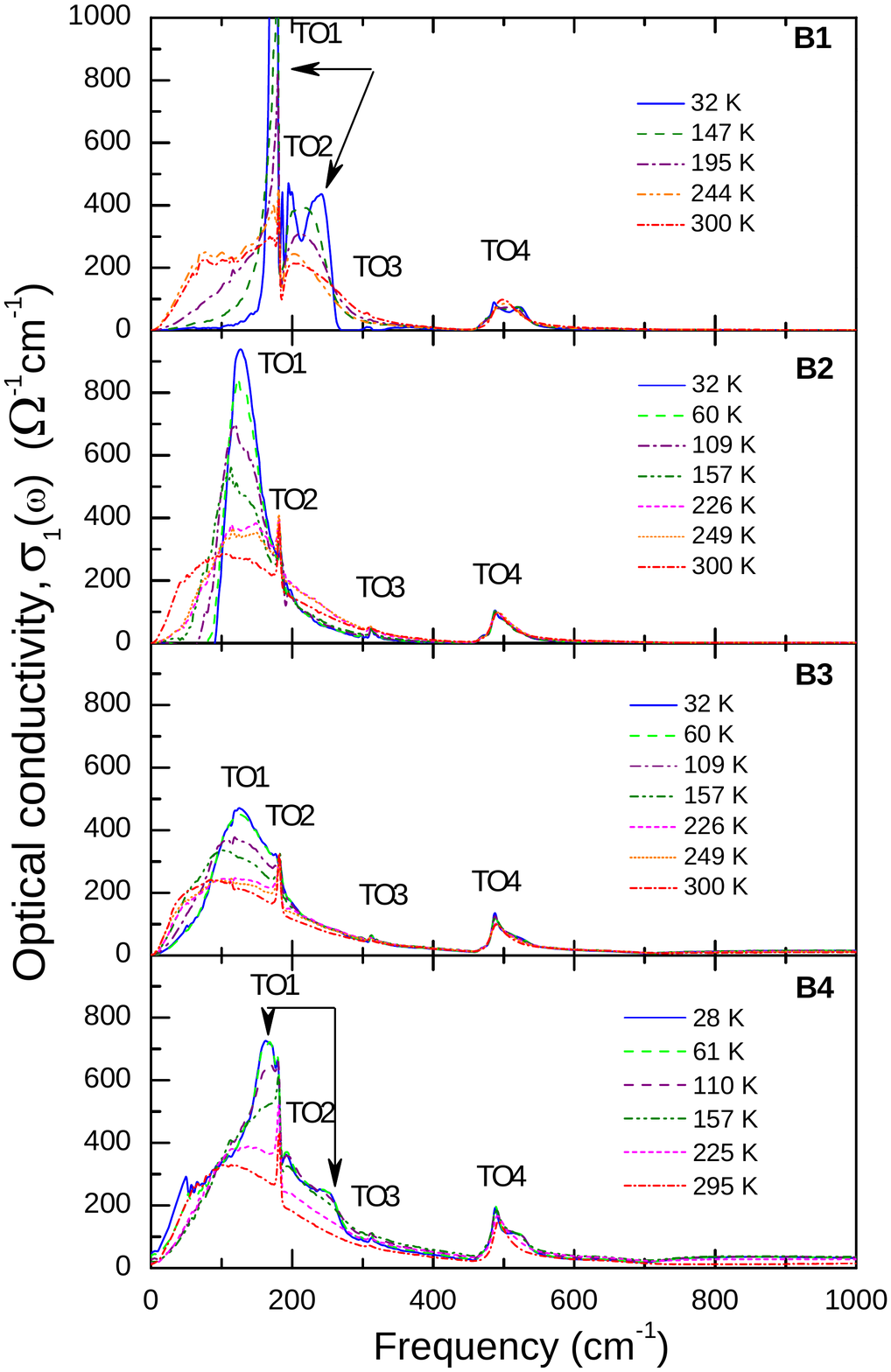}
         \vspace*{-1.0cm}%
        \end{center}
\caption{(color online) Optical conductivity of \btd\ samples extracted from reflectance by using the Kramers-Kronig analysis.} \label{fig5}
\end{figure}

Temperature evolution of the optical conductivity of \btd\ samples in the 0 -- 1000 cm$^{-1}$ range is shown in Figure \ref{fig5}. For cubic \bt, the group theory predicts three triply degenerate IR active modes of {\it F$_{1u}$} symmetry and one triply degenerate IR silent mode of {\it F$_{2u}$} symmetry\cite{nakagawa67}. The three {\it F$_{1u}$} modes in a frequency order from low to high are identified as the Slater mode (Ti-O bond length modulation), the Last mode that involves translational motion of the TiO$_6$ octahedron against Ba atoms, and the Axe mode (Ti-O-Ti bond angle modulation). The {\it F$_{2u}$} mode associated with torsional motion of oxygen ions becomes IR and Raman active in the low-symmetry phases.

According to the literature, upon cooling the Slater mode undergoes softening in the cubic phase. It saturates at $\omega_{TO1} \approx$ 31 cm$^{-1}$ upon approaching $T_C$ = 403 K.\cite{vogt82} According to the displacive model of the phase transition, it is this (soft) mode that is responsible for the ferroelectric instability in \bt.\cite{barker66,cochran60} On the other hand, the Last ($\omega_{TO2} \approx$ 180 cm$^{-1}$) and the Axe ($\omega_{TO4} \approx$ 498 cm$^{-1}$) modes are nearly $T$-independent in a wide temperature range. Below $T_C$,
the Last and Axe modes develop fine structure in the low-symmetry phases but their central frequencies remain nearly constant.\cite{comment01} The $E$ mode that originates from the splitting of the {\it F$_{2u}$} cubic silent mode shows a very weak intensity. Its frequency remains $T$-independent at $\omega_{TO3} \approx$ 310 cm$^{-1}$ in good agreement with the Raman data.\cite{perry65,scalabrin77}

Except for the TO1 mode, doping with charge carriers does not have a significant effect on the TO phonon frequencies in \btd\ (Figure \ref{fig5}). Colossal decrease by ca. 110 cm$^{-1}$ in the TO1 soft mode frequency is found for semiconducting B2 and B3 samples as compared with undoped B1 sample. This effect is opposite to the case of the \std\ and Nb-doped \st\ crystals where the soft mode frequency increases with carrier density in a wide concentration range.\cite{bauerle80,mechelen08} In the rhombohedral phase, $T <$ 195 K,  metallic B4 sample shows splitting of the soft mode into the low- and high-frequency components. The high-frequency component detected as a shoulder in the $\sigma_1(\omega)$ plot at $\omega \approx 242$ cm$^{-1}$ recovers to that of the undoped \bt\ at $\omega_{TO1} \approx$ 241 cm$^{-1}$. The strongly asymmetric low-frequency component of the soft mode has a peak at $\omega \approx$ 166 cm$^{-1}$ which is ca. 40 cm$^{-1}$ higher than $\omega_{TO1}$ of semiconducting B2 and B3 samples (Figure \ref{fig5}).

\begin{figure}[t!]
        \begin{center}
         \vspace*{-1.5 cm}%
               \leavevmode
                \includegraphics[origin=c, angle=0, width=8 cm, clip]{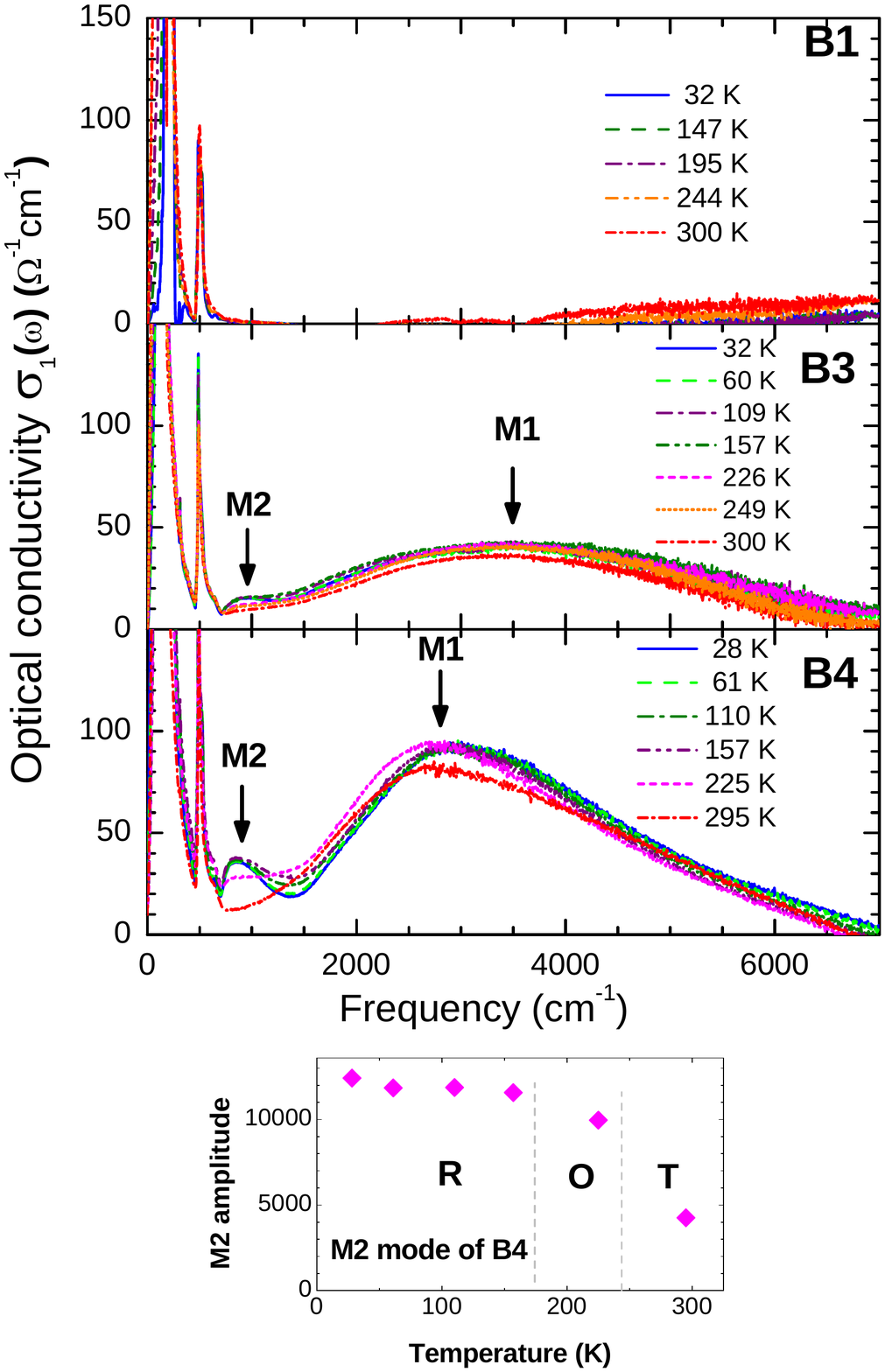}
         \vspace*{-0.5 cm}%
        \end{center}
\caption{(color online) Optical conductivity of B1, B3, and B4 samples in the 0 -- 7000 cm$^{-1}$ frequency range. In the lowest panel we show the temperature dependent M2 amplitude of B4 sample.} \label{fig6}
\end{figure}

Figure \ref{fig6} displays a mid-infrared (MIR) absorption band at 700 -- 7000 cm$^{-1}$ which develops in the heavily doped B3 and B4 samples. As doping increases, the MIR band gets stronger and better-resolved. At room temperature, the MIR band is characterized by a broad peak denoted M1 in Fig. \ref{fig6}. The peak position shifts down from 3500 cm$^{-1}$ for B3 sample to 2600 cm$^{-1}$ for more reduced metallic B4 sample. This doping dependence of the MIR absorption band is similar to the earlier report on reduced BaTiO$_{3-x}$ epitaxial thin films\cite{zhao00} and La$_{2-x}$Sr$_x$CuO$_4$ cuprates.\cite{uchida91} On cooling from 300 K, M1 peak shifts slightly to the higher frequency and a new sharper and smaller peak, denoted M2 in the figure, develops at the lower frequency. Intensity of the M2 peak centered at 890 cm$^{-1}$ (for $n = 3.1 \times 10^{19}$ cm$^{-3}$) and at 840 cm$^{-1}$ (for metallic sample) increases with decreasing temperature and tends to saturate in the rhombohedral phase; we estimate the area under the M2 mode by fitting the $\sigma_1(\omega)$ of our B4 sample with Lorentzian functions (not shown) and display the resulting temperature dependent amplitude in the lowest panel of Fig. \ref{fig6}. The M2 amplitude tends to saturate in the Rhombohedral phase. The temperature dependent amplitude suggests that the appearance of the M2 absorption band correlates with the structural phase transitions.

%
%

\section{Discussion}

\subsection{Soft mode behavior}

\begin{figure}[t!]
        \begin{center}
         \vspace*{-1.5 cm}%
               \leavevmode
                \includegraphics[origin=c, angle=0, width=9 cm, clip]{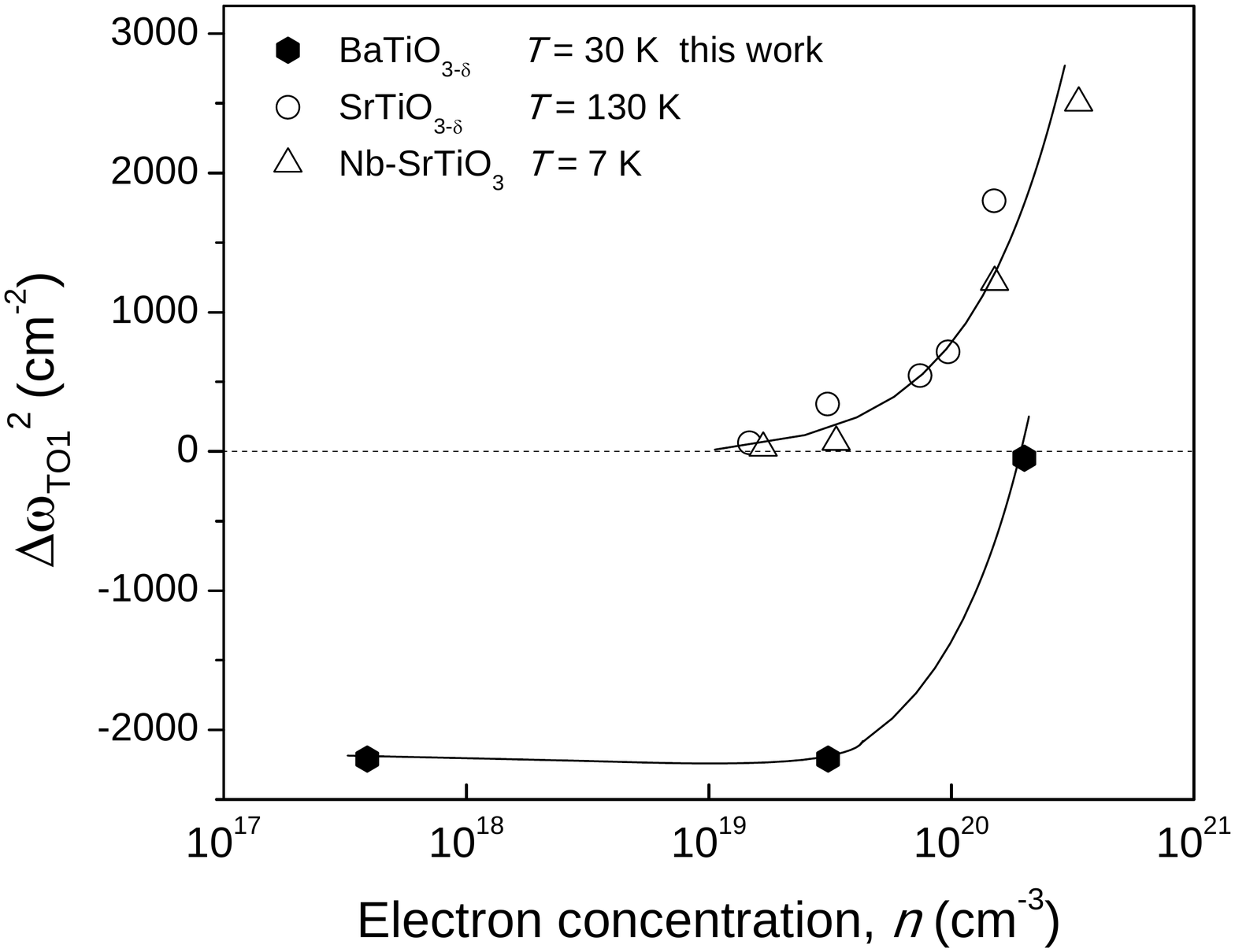}
         \vspace*{-1.0cm}%
        \end{center}
\caption{Electron concentration dependence of the shift of the soft mode squared. The data for Nb-doped and reduced \st\ were adopted from reported work by van Mechelen {\it et al.}\cite{mechelen08} and B\"{a}uerle {\it et al.}\cite{bauerle80}, respectively. The lines are the guides to the eye.} \label{fig7}
\end{figure}

Unusually strong and somewhat unexpected doping dependence of the soft phonon mode in \btd\ has been found in this work. Electron concentration dependence of the shift of the soft mode frequency squared, $\Delta\omega_{TO1}^2$, is shown in Figure \ref{fig7}. In contrast to V$_{\mbox{o}}$-doped and Nb-doped \st\ samples that show hardening of the soft mode frequency with doping concentration, the frequency of the soft mode in \btd\ decreases significantly in the semiconducting samples. However, in metallic \btd, the $\omega_{TO1}$ nearly recovers to the value found in undoped \bt.

The origin of the soft mode hardening in \std\ has been discussed by B\"{a}uerle et al.\cite{bauerle80} Based on the formalism proposed by Trunov {\it et al.}\cite{trunov74} it was demonstrated that the role of the free carrier concentration in the soft mode hardening is negligible. The authors, therefore, have attributed the soft mode dependence in \std\ to the change in the long-range dipole-dipole interaction and local potential due to oxygen vacancies.\cite{bauerle80} This interpretation, however, cannot explain very similar hardening of the soft mode in Nb-doped \st\ samples that are free from oxygen vacancies\cite{gervais93,mechelen08} (see Fig. \ref{fig7}).

Two types of oxygen atoms can be distinguished in tetragonal BaTiO$_{3}$: one in the Ti-O-Ti chains along the c-axis [O(1)] and the other in the Ti-O-Ti orthogonal chains in the ab-plane (along a- and b-axis) [O(2)]. Therefore, one may expect two types of oxygen vacancy defects in \bt, i.e., V$_{\mbox{o}}^{c}$ and V$_{\mbox{o}}^{ab}$. First-principles calculations of oxygen vacancies in tetragonal PbTiO$_{3}$ reveal that the energy difference between V$_{\mbox{o}}^{c}$ and V$_{\mbox{o}}^{ab}$ is around 0.3 eV with V$_{\mbox{o}}^{c}$ being more energetically stable.\cite{park98} Calculation of atomic displacements closest to the V$_{\mbox{o}}^{c}$ indicate that Ti ions due to Coulomb repulsion are displaced away from the V$_{\mbox{o}}^{c}$ defect. Therefore, according to a simple model, the shorter Ti-O distance should increase the force constants of remaining Ti-O bonds causing enhancement of the Slater phonon frequency.\cite{last57} The Slater mode dependence in insulating \btd\ shows, however, an opposite trend.

We conjecture that the remarkable softening of the TO Slater mode in the insulating \btd\ is associated with the strong electron-lattice interaction and possible formation of the small polaron.\cite{lenjer02,kolodiazhnyi06} Kostur and Allen\cite{kostur97} indicate that the small polaron formation will result in renormalization of both electron energy and the energy of the local vibrations near the localized electron. Based on the 3D model of the small polaron in Ba$_{1-x}$K$_x$BiO$_3$, the authors find significant softening of the Bi-O$_6$ breathing phonon
mode.\cite{kostur97} In addition, if the majority of the small polarons in \btd\ are isolated Ti$^{3+}$ defects, as indicated by EPR, the simple geometric considerations suggest that the magnitude of their off-center diagonal displacement will be reduced as compared to remaining Ti$^{4+}$ ions. Since the vibrational frequency is mostly determined by the shortest Ti-O distance, the overall effect of the Ti$^{3+}$ defects would be to further decrease the Slater mode frequency.

\subsection{Mid infrared band}

Interpretation of the MIR absorption band in n-type \bt\ has been a topic of significant interest and still remains
controversial.\cite{berglund67,gerthsen65,gerthsen65a,bursian71,ihrig76} Early reports \cite{gerthsen65,gerthsen65a} have attributed the M1-like absorption band in Figure \ref{fig6} to optical excitations of small polarons based on satisfactory agreement between optical (\emph{E}$_{\mathrm{op}}$ = 500 -- 600 meV) and thermal (\emph{E}$_{\mathrm{th}}$ = 100 meV) activation energies, that for small polarons should satisfy \emph{E}$_{\mathrm{op}}$ = 4\emph{E}$_{\mathrm{th}}$. This interpretation has been later questioned by Berglund and Braun \cite{berglund67} who have assigned the M1 absorption to the oxygen vacancy related defects based on the M1 anisotropy and temperature dependence in the tetragonal phase. Theoretical studies reveal that the ground state of the singly ionized oxygen vacancy, V$_{\mbox{o}}^{\cdot}$, involves a transfer of electron density to the $3d_{3z^2-r^2}$ orbital of the two neighbor axial Ti ions.\cite{donnerberg00,solovyev08} Clustering of the two oxygen vacancies in the form of the axial [V$_{\mbox{o}}^{\cdot\cdot}$-Ti$^{2+}$-V$_{\mbox{o}}^{\cdot\cdot}$]$^{\cdot\cdot}$ complex defect with 2 electrons localized on the Ti $3d_{3z^2-r^2}$ orbital forming an in-gap electronic state at 0.6 eV below conduction band minimum has been predicted by Cuong et al.\cite{cuong07}

In a relevant study on Nb-doped \st, Mechelen et al. have attributed MIR absorption to the multiphonon sidebands due to the incoherent electron-phonon coupling.\cite{mechelen08} They found that the missing Drude spectral weight is nearly recovered in the MIR range. In agreement with a large polaron formalism developed by Devreese et al.,\cite{devreese72} the MIR band shows additional low-energy peak upon cooling. The intensity of the MIR band in our \btd\ samples is comparable to that of the Nb-doped \st\ at similar electron density.\cite{mechelen08} Both data sets also show similar red shift of the MIR peak frequency with electron doping. Finally, we observe a second narrower MIR band (M2) that emerges upon cooling on the low-energy side. Our data, however, shows a slight blue shift of the M1 MIR peak upon cooling in agreement with a small polaron scenario.\cite{emin93,fratini06} Similar temperature dependence of the MIR band has been reported in semiconducting Nd$_{1-x}$TiO$_{3}$.\cite{yang06}

Although our MIR data do not contradict the small polaron interpretation, we would like to exercise caution here. Over the course of this study it became obvious that in order to clearly separate the possible small polaron and oxygen vacancy contributions to MIR band, an additional optical study on La- or Nb-doped \bt\ single crystals is required.

\section{Conclusions}

Optical properties of \btd\ single crystals reveal that the symmetry-induced splitting of the fundamental phonon modes survive for all the samples studied including metallic one. This provides further evidence of the `ferroelectric metal' ground state in metallic \btd.

Upon doping with oxygen vacancies we found strong renormalization of the low-energy soft phonon mode, $\omega_{TO1}$, in \btd\ which is qualitatively different from V$_{\mbox{o}}$-doped or Nb-doped \st. In particular, the low-temperature $\omega_{TO1}$ in the insulating \btd\ samples is ca. 28 $\%$ lower than that of undoped \bt, in very good agreement with electron-phonon coupling model of Kostur and Allen.\cite{kostur97} In metallic \btd, due to the free electron screening, the soft mode energy nearly recovers to the value of undoped \bt.

Mid-infrared absorption appears in reduced \btd\ samples and increases with electron concentration. Upon cooling the MIR band is split into two bands which is similar to the behavior of V$_{\mbox{o}}$-doped or Nb-doped \st. We find a correlation between appearance of the M2 band and the structural phase transition in the system. Both M1 and M2 bands soften and get stronger as the V$_{\mbox{o}}$ concentration increases. Based on the MIR temperature dependence we tentatively assign the M1 and M2 bands to optical excitation of small polarons. In order to confirm this assignment, more studies on La- or Nb-doped \bt\ crystals that are free from oxygen vacancies are needed.

\begin{acknowledgments}
The authors thank Thomas Timusk for supplying useful comments and supporting this work. This work has been supported by the Natural Science and Engineering Research Council of Canada and the Canadian Institute for Advanced Research. J. Hwang acknowledges financial support from from the National Research Foundation of Korea (NRFK No. 20100008552). The work of T.K. was funded by Grant-in-Aid for Scientific Research C Grant $\#$ 21560025 provided by MEXT Japan.
\end{acknowledgments}

%
%

\end{document}